\documentclass{emulateapj}

\usepackage{color}
\newcommand{\DUEDATE}{May 1, 2010}
\newcommand{\SNPCC}{{\tt SNPhotCC}}

\newcommand{\unc}{uncertainty}
\newcommand{\uncs}{uncertainties}
\newcommand{\id}{identification}
\newcommand{\phot}{photometric}

\newcommand{\specy}{spectroscopically}
\newcommand{\spec}{spectroscopic}

\newcommand{\wwwPanSTARRS}{\tt http://pan-starrs.ifa.hawaii.edu/public}
\newcommand{\wwwNugent}{\tt http://supernova.lbl.gov/nugent/nugent\_templates.html}

\newcommand{\rspeclimit}{21.5}
\newcommand{\ispeclimit}{23.5}
\newcommand{\wwwdownload}{\tt www.hep.anl.gov/SNchallenge}
\newcommand{\photoz}{photo-$z$}

\newcommand{\NIaTOT}{{\cal N}_{\rm Ia}^{\rm TOT}}
\newcommand{\NIatrue}{N_{\rm Ia}^{\rm true}}
\newcommand{\NIafalse}{N_{\rm Ia}^{\rm false}}
\newcommand{\wIafalse}{W_{\rm Ia}^{\rm false}}
\newcommand{\effIa}{\epsilon_{\rm Ia}}
\newcommand{\effspec}{\epsilon_{\rm spec}}
\newcommand{\FoMIa}{{\cal C}_{\rm FoM-Ia}}

\begin{document}
\title{Supernova Photometric Classification Challenge}

\submitted{Challenge Released on Jan 29, 2010. Last update: \today}
%%\submitted{}
%% \email{kessler@kicp.uchicago.edu}

% ===========================================================
%     author list for SDSS/SNLS/LSST photoz paper  
% ===========================================================
% institution nick-names

\newcommand{\NUMUCASTRO}{1}   % U.C astro
\newcommand{\NUMKICP}{2}      % U.Chicago KICP
\newcommand{\NUMUCB}{3}       % U.Colorado, Boulder
\newcommand{\NUMRUTGERS}{4}   % Rutgers University
\newcommand{\NUMANL}{5}        % Argonne National Lab

% ====================================================
% names

\author{
Richard~Kessler,\altaffilmark{\NUMUCASTRO,\NUMKICP}
Alex~Conley,\altaffilmark{\NUMUCB}
Saurabh~Jha,\altaffilmark{\NUMRUTGERS}
Stephen~Kuhlmann\altaffilmark{\NUMANL}
} % end authorlist

% ===============================================
%       INSTITUTIONS
% ===============================================

\altaffiltext{\NUMUCASTRO}{
  Department of Astronomy and Astrophysics,
   The University of Chicago, 5640 South Ellis Avenue, Chicago, IL 60637
}

\altaffiltext{\NUMKICP}{
  Kavli Institute for Cosmological Physics, 
   The University of Chicago, 5640 South Ellis Avenue Chicago, IL 60637
}

\altaffiltext{\NUMUCB}{
Center for Astrophysics and Space Astronomy,
University of Colorado, Boulder, CO, 80309-0389, USA
}

\altaffiltext{\NUMRUTGERS}{
Department of Physics and Astronomy, 
Rutgers University, 136 Frelinghuysen Road, Piscataway, NJ 08854
}

\altaffiltext{\NUMANL}{
Argonne National Laboratory, 9700 S. Cass Avenue, Lemont, IL 60437
}

% =========== END ==============

% ===========================
\begin{abstract}
We have publicly released a blinded mix of simulated SNe,
with types (Ia, Ib, Ic, II) selected in proportion to 
their expected rate. The simulation is realized in the 
$griz$ filters of the Dark Energy Survey (DES)
with realistic observing conditions 
(sky noise, point spread function and atmospheric transparency) 
based on years of recorded conditions at the DES site.
Simulations of non-Ia type SNe are based on 
\specy\ confirmed light curves that include 
{\it unpublished} non-Ia samples donated from the 
Carnegie Supernova Project (CSP), the
Supernova Legacy Survey (SNLS), and the
Sloan Digital Sky Survey-II (SDSS--II).
We challenge scientists to run their classification
algorithms and report a type for each SN.
A \specy\ confirmed subset is provided for training.
The goals of this challenge are to 
(1) learn the relative strengths and weaknesses of the different 
classification algorithms, 
(2) use the results to improve classification algorithms, and
(3) understand what \specy\ confirmed sub-sets are needed to
properly train these algorithms.
The challenge is available at {\wwwdownload},
and the due date for classifications is {\DUEDATE}.
\keywords{supernova light curve fitting and classification}
\end{abstract}

% ##################################
 \section{Motivation}
 \label{sec:intro}
% ##################################

To explore the expansion history of the universe, 
increasingly large samples of high quality SNe~Ia 
light curves are being used to measure luminosity 
distances as a function of redshift.  
With increasing sample sizes, there are 
not nearly enough resources to \specy\ confirm each SN.
Currently, the world's largest samples are from the 
Supernova Legacy Survey (SNLS: \cite{Astier06}) and the
Sloan Digital Sky Survey-II (SDSS-II: \citet{Frieman07}),
each with more than 1000 SNe~Ia, 
yet less than half of their SNe are \specy\ confirmed.
The numbers of SNe are expected to increase dramatically
in the coming decade:
thousands for the Dark Energy Survey (DES: \citet{DES-moriond2009}) 
and a few hundred thousand for the 
Panoramic Survey Telescope and Rapid Response System
(Pan-STARRS)\footnote{\wwwPanSTARRS}
and the Large Synoptic Survey Telescope
(LSST: \citet{Ivezic_08,LSSTSB09}).
Since only a small fraction of these SNe will be \specy\ confirmed,
\phot\ \id\ is crucial to fully exploit these large samples.

In the discovery phase of accelerated cosmological expansion,
results were based on tens of high-redshift SNe~Ia,
and some samples included a significant fraction of events 
that that were not classified from a spectrum
\citep{Riess_1998,Perl_1999,Tonry_2003,Riess_2004}.
While human judgment played a significant role in classifying
these SNe without a spectrum,
more formal methods of \phot\ classification have been
developed over the past decade:
\citet{Poz2002,Dahlen2002,Sulli2006,Johnson2006,Poz2007,Kuz2007,Rodney2009}.
Some of these methods have been used to select candidates
for \spec\ follow-up observations, but these methods have not
been used to select a significant \phot\ SN~Ia sample for a 
Hubble diagram analysis.
In short, cosmological parameter estimates
from much larger and recent surveys are based
solely on \specy\ confirmed SNe~Ia
(SNLS: \citet{Astier06}, ESSENCE: \citet{WV07},
CSP: \citet{Freedman_2009}, SDSS-II: \citet{K09}).

The main reason for the current reliance on \spec\ \id\
is that vastly increased \spec\ resources have been used in 
these more recent surveys. 
In spite of these increased resources,
more than half of the discovered SNe do not have a spectrum 
and therefore \phot\ methods will eventually be needed
to classify the majority of the SNe.
There are two difficulties limiting the application 
of \phot\ classification.
First is the lack of adequate non-Ia data for training algorithms.
Many classification algorithms were developed using 
non-Ia templates\footnote{\wwwNugent}
constructed from averaging and interpolating a limited 
amount of \specy\ confirmed non-Ia data,
and therefore the impact of the non-Ia diversity has 
not been well studied.
The second difficulty is that there is no standard 
testing procedure, and therefore it is not
clear which classification methods work best.

To aid in the transition to using \phot\ SN-classification,
we have released a public ``SN  Photometric Classification Challenge'' 
to the community, hereafter called {\SNPCC}.
The \SNPCC\ consists of a blinded mix of simulated SNe,
with types (Ia, Ib, Ic, II) selected in proportion to 
their expected rate.
The challenge is for scientists to run their classification
algorithms and report a type for each SN.
A \specy\ confirmed sub-set is provided so that algorithms
can be tuned with a realistic training set.
The goals of this challenge are to 
(1) learn the relative strengths and weaknesses of the different 
classification algorithms, 
(2) use the \SNPCC\ results to improve the algorithms, and
(3) understand what \specy\ confirmed sub-sets are needed to
properly train these algorithms.

To address the paucity of non-Ia data, the CSP, SNLS and SDSS-II
have contributed {\it unpublished} 
\specy\ confirmed non-Ia light curves.
These data are high-quality multi-band light curves,
not just junk that nobody cares about, 
and therefore we are grateful to the donating collaborations.
This non-Ia sample is likely to undersample the 
potential variety in the upcoming surveys like DES and LSST, 
but we anticipate that this challenge will
be a useful step away from the overly-simplistic
studies that have relied on a handful of non-Ia templates.

The outline of this release-note is as follows.
A description of the simulation is given in 
\S\ref{sec:sim}, and instructions for participants are in
\S\ref{sec:challenge}.
Comments on the evaluations and posting of results are
given in \S\ref{sec:eval}.

% ##################################
 \section{The Simulation}
 \label{sec:sim}
% ##################################

The simulation is realized in the $griz$ filters of the 
Dark Energy Survey (DES). The sky-noise, point-spread function
and atmospheric transparency are evaluated in each filter
based on years of observational data from the ESSENCE project
at the Cerro Tololo Inter-American Observatory (CTIO).
For the five SN fields (3 sq deg each), the cadence is based
on allocating 10\% of the DES photometric observing time
and most of the non-photometric time.
The cadence used in this publicly available simulation was 
generated by the Supernova Working Group within the 
DES collaboration.\footnote{Although two of us (RK \& SK) 
are members of the DES,
we have not included other DES colleagues in any discussions
about this challenge, and we have made our best efforts to 
prevent our DES collaborators from obtaining additional 
information beyond that contained in this note.
} % end footnote
Since the DES plans to collect data during 5 months of the year,
incomplete light curves from temporal edge effects are included;
i.e., the simulated explosion times extend well before the start 
of each survey season, and extend well beyond the end of the season.

Simulated SNe~Ia are  based on models empirically derived from data.
In addition to the model parameters, we have
applied tweaks to simulate the anomalous Hubble scatter.
While these tweaks are invented ad-hoc, they have not been
ruled out with current observations.
Simulated non-Ia SNe are based on observed multi-color
light curves (from CSP, SNLS, and SDSS)
that have been smoothed in each passband,
and then K-corrected to the appropriate redshift and filters.

% When a measured non-Ia light curve is K-corrected to a  
% smaller redshift, the $z$-band prediction is unreliable 
% and is therefore excluded from the light curve;
% this affects only a small fraction of the simulated SNe.

A \specy\ confirmed subset is based on observations
on a 4~meter class telescope with a limiting $r$-band
magnitude of $\rspeclimit$, and an 8~meter class telescope
with a limiting $i$-band magnitude of $\ispeclimit$.
The subset is randomly selected, and the number of 
\specy\ confirmed SNe ($\sim 1000$) corresponds to the 
combined resources of the SNLS \& SDSS--II surveys.
While this number of \spec\ identifications may be
optimistic, this allows for further study on how the
training quality depends on the size of the \spec\ sample.

For the challenge that includes the host-galaxy 
photometric redshift, the \photoz\ estimates are based 
on simulated galaxies (for DES) analyzed with the
methods in \citet{Oy08a,Oy08b}.
The average host-galaxy \photoz\ resolution is 0.03.

Two simple selection criteria have been applied.
First, each object must have at least one observation 
with a signal to noise ratio (S/N) above 5 (in any filter).
Second, there must be at least 5 observations after explosion,
and there is no S/N requirement on these observations.
These requirements are relatively loose because part of 
the challenge is to determine the optimal selection criteria.
The total number of simulated SNe that satisfy these loose
selection requirements is $2\times 10^4$,
and corresponds to the 5 seasons planned for the DES.

% ##################################
 \section{Taking the Challenge}
 \label{sec:challenge}
% ##################################

Two independent challenges have been generated:
one with a host-galaxy \photoz, and another without
any redshift information.
In addition to these challenges based on the entire light curve, 
there is also an early-epoch challenge based on the first six 
observations (in any filter) with ${\rm S/N} > 4$. 
On the night of the sixth observation, all observations 
made this night are included.
Among the four challenges available, 
you may take any of them or all of them.

The simulated light curves can be downloaded from the
\SNPCC\ website.\footnote{\wwwdownload}
The filter response functions are given in the files
{\tt DES\_[griz].dat}.
The file with the ``{\tt .LIST}'' suffix provides
a list of all data files to analyze.
The data files are self-documented and visual
inspection should be adequate for preparing 
a parsing algorithm. The calibrated fluxes
are defined as
\begin{equation}
  {\rm\tt FLUXCAL} = 10^{(-0.4\cdot m +11)} + {\rm noise}
  \label{eq:fluxcal}
\end{equation}
where $m$ is the modeled AB-magnitude of the SN,
and the noise contributions\footnote{The noise has been
scaled from photoelectrons into {\tt FLUXCAL} units.}
include Poisson fluctuations, sky noise, and CCD noise.
The observed magnitudes are not provided because they are 
not defined when noise fluctuations result in a negative flux;
for fitting, we recommend translating model-magnitudes 
into fluxes as defined in Eq.~\ref{eq:fluxcal}.

For tuning your algorithms, the \specy\ confirmed sub-sample
is identified by the {\tt SNTYPE} keyword (see Table~\ref{tb:types}),
and the corresponding redshift is given by the 
{\tt REDSHIFT\_SPEC} keyword. For the majority of SNe
that do not have \spec\ identification, 
the type and \spec\ redshift are set to $-9$.
For the host-galaxy \photoz\ sample, the \photoz\ is given
by the {\tt HOST\_GALAXY\_PHOTO-Z} keyword.
For the early-epoch challenge, process only the observations
that appear before the ``{\tt DETECTION:}'' keyword.

A valid challenge submission must contain three items:
(1) an answer list containing the type for each SN, 
(2) a brief description of your method, and
(3) an estimate of the CPU resources.
For a group effort, a team name is recommended.
These submission items are discussed below in more detail.

For each challenge that you participate in,
your answer list must contain four columns: 
\begin{verbatim}
  SNID   TYPE   PHOTOZ   PHOTOZ_ERROR
\end{verbatim}
where 
\begin{itemize}
  \item {\tt SNID} is the SN integer id
  \item {\tt TYPE} is the integer SN-type code returned by
        your classifier (see Table~\ref{tb:types}).
 	You can report either a general type (1,2,3 for Ia,II,Ibc),
	or a specific sub-type.
  \item {\tt PHOTOZ} is \photoz\ value returned by your classifier.
  \item {\tt PHOTOZ\_ERROR} is the uncertainty 
\end{itemize}
If your code does not return a useful \photoz\ value,
just set $-9$ in the last two columns.  
A valid answer list must contain entries in all
four columns and for each SN; invalid answer files
will be returned. In addition to the answer file,
please provide a brief description of your technique.
A reference to either a refereed journal article
or arXiv posting is adequate, but please describe 
any modifications from the referenced article.
Finally, include the processing time,
the number of light curves analyzed 
(i.e, that are not rejected by selection cuts)
and a description of your computing processor hardware.

In addition to thinking about your classification algorithm,
you should also think about appropriate selection cuts
to reject SNe that are difficult to classify.
Set the SN type to $-1$ for rejected SNe. 
As described in \S\ref{sec:eval}, our evaluation generally
penalizes incorrect classifications more than it penalizes 
the loss from selection cuts.

To maximize the utility of this challenge, 
please respect the following guidelines.
While you can use the \specy\ confirmed subset to
train your algorithms, please use your program to
report classifications from this subset;
i.e, do not just report the \spec\ SN type.
A useful diagnostic in the evaluation will be
to compare the classification performance 
from the training subset to that from the rest of the sample.
In a similar spirit, do {\it not} use the \spec\ redshift 
({\tt REDSHIFT\_SPEC}) to report classifications.
Finally, for the early-epoch challenge, use only the \specy\ 
confirmed sub-sample for tuning your algorithms;
i.e., do {\it not} use the full set of (unconfirmed) light curves.

Don't hesitate to report problems or suggestions,
including methods for evaluation.
Missing information and updates will be appended to
\S\ref{sec:updates} and re-posted to the arXiv.
You should periodically check this arXiv posting
for updates.

\bigskip
Finally, the due date is {\bf \DUEDATE}.

\begin{table}  %  [hb]
\caption{Integer codes for SN types.}
\begin{center}
\begin{tabular}{l | l }
\tableline\tableline
                &  integer   \\
 SN-type        &  code      \\
\tableline % ------------------------------------------------
  Ia                  & 1    \\
  II (IIn, IIP, IIL)  & 2 (21, 22, 23)  \\
  Ibc (Ib, Ic)        & 3 (32, 33)   \\
  other        & 66   \\
  rejected     & $-1$ \\
\tableline  % ------------------------------------------
\end{tabular}
\end{center}
  \label{tb:types}
\end{table}

% ##################################
 \section{Posting \& Evaluating the Challenge Results}
 \label{sec:eval}
% ##################################

Classification results from the participants will
be posted publicly along with our initial evaluations
and the answer key.
Anyone can therefore evaluate the algorithms
using their choice of figure-of-merit (FoM).
We will also provide additional information about the 
simulation strategy, along with details for each simulated SN.
For non-Ia type SNe based on K-correcting unpublished
light curves, the level of detail that we release
will be determined solely by the donating collaborations.
Shortly before posting the answer key, we will ask
the donating collaborations for instructions on 
what details can be released.

We finish with a discussion of ideas on how to evaluate the results.
Ideally, we would like to assign a single number (FoM) 
for each algorithm. To make more refined
comparisons, the FoM can be tabulated as a function 
of redshift or any other variable of interest.

We begin the discussion by considering the FoM for
a Ia rate measurement based on \phot\ \id.
After selection requirements have been applied,
let $\NIatrue$ be the number of correctly typed SNe~Ia,
and $\NIafalse$ be the number of non-Ia that are 
incorrectly typed as an SN~Ia. A simple classification FoM 
is the square of the signal-to-noise ratio (S/N) divided
by the total number of SNe~Ia ($\NIaTOT$) before selection cuts,
\begin{eqnarray}
    \FoMIa &  \equiv & 
     \frac{1}{\NIaTOT} \times 
     \frac{({\NIatrue})^2}{\NIatrue + \wIafalse\NIafalse}  
          \nonumber  \\
%%            &     &    \\
            &     =   & \effIa \times
                       [ \NIatrue/(\NIatrue + \wIafalse\NIafalse) ]~,
   \label{eq:FoM1}
\end{eqnarray}
where $\wIafalse$ is the false-tag weight (penalty factor)
described below,
$\effIa$ is the SN~Ia efficiency that includes
both selection and typing requirements,
and $\NIatrue = \effIa\NIaTOT$.
Since $\NIaTOT$ is a constant that is independent of the analysis, 
we have divided out this term so that $0 \le \FoMIa \le 1$, 
with $\FoMIa = 1$ corresponding to the theoretically optimal analysis.

The FoM in Eq.~\ref{eq:FoM1} is the product of two terms.
The first term is the efficiency for selecting and classifying 
type Ia SNe, and the second term is the Ia purity (when $\wIafalse=1$), 
the fraction of classified Ia that really are SNe~Ia.
In the ideal case where the average of $\NIafalse$ is 
perfectly determined, $\wIafalse = 1$ and the naive 
statistical \unc\ is the only contribution to the FoM.
In practice, \uncs\ in determining the false-tag rate lead to
$\wIafalse > 1$. For example, suppose that $\NIafalse$ is 
scaled from a \specy\ confirmed subset containing a 
fraction ($\effspec$) of the total number of SNe;
in this case, $\wIafalse = 1+1/\effspec$ is much larger
than 1 if the \spec\ subset is small.
It may be possible to reduce $\wIafalse$ using other methods 
to determine $\NIafalse$, such as fitting the tails in the 
distance-modulus residuals. 
For SN-cosmology applications, a proper determination of 
$\wIafalse$ is beyond the scope of this classification challenge, 
but suggestions are welcome on setting an appropriate
value for the evaluations.

Next we illustrate the FoM with a numerical example
in which the false-tag rate is determined from a \spec\
sub-sample with $\effspec = 0.2$, and $\wIafalse = 6$.
Consider a sample with 50\% type Ia
and 50\% non-Ia, and $\effspec = 0.2$.
Assume that the classification algorithm correctly 
identifies half of the SNe,  while for the other half
the classification works so poorly that it is equivalent
to making random guesses with a 50\% probability of 
guessing correctly. 
If the ambiguous half is rejected, then $\effIa = 0.5$, 
the purity term is 100\% (since $\NIafalse = 0$), 
and $\FoMIa = 0.5$.
Now consider an analysis strategy without selection requirements.
The efficiency term increases to $\effIa = 75\%$ since 25\%
of the SNe~Ia are rejected by incorrect classifications.
However, since the false-classification rate increases to
$\NIafalse/\NIatrue = 1/3$
%  NIatrue = 25 + .5*25 = 37.5; NIafalse = .5*.5*50 = 12.5
the purity term drops to $1/(1+6\cdot 1/3) = 1/3$
and the net FoM drops to $\FoMIa = 1/4$.
An algorithm that simply makes a random guess on all SNe
results in $\FoMIa = 1/14$.
% Eff=.5;  NIatrue=25;  NIafalse=25;  purity = 1/(1+6) = 1/7
The point of this exercise is to illustrate the importance
of selection criteria, and that forcing a classification
on every SN candidate is not necessarily the optimal strategy.

% ##################################
%% \clearpage
 \section{Post-Release Updates}
 \label{sec:updates}
% ##################################

\begin{itemize}
  \item {\bf February 7, 2010:} for the \specy\ confirmed subset, 
	sub-types are given as indicated in Table~\ref{tb:types}. 
	Participants can either report a general classification 
	(i.e., 1,2,3 $\to$ Ia,II,Ibc) or report a 
	specific sub-type (e.g., IIn, Ic, etc.). 
	Download the updated challenge data files only if you 
	need the sub-types.
  \item {\bf March 14, 2010:} Fixed bug in which about 1\% of the SNe
	have pathological late-time magnitudes. Download data files
	after date-stamp above.
  \item {\bf March 24, 2010:} Fixed bug in which a few dozen non-Ia SNe        
	have pathological magnitudes at all epochs.
  \item {\bf April 13, 2010:} Fixed two bugs related to type II SNe.
        First, the wrong redshift was mistakenly used for one of the 
	observed IIP, resulting in a 2~mag overestimate
	of its brightness. 
	Second, for another type II SN the absolute mag was 
	mistakenly set 0.3 mag too bright.
	While the generated fraction of these buggy SNe was small, 
	their contribution to the challenge sample after requiring 
	S/N$>5$ was relatively large; 
	therefore the updated sample has $\sim 1400$ fewer SNe.
 \item {\bf April 27, 2010:} No bug-fixes, but we have decided to
    	to fix $\wIafalse = 3$ for the $\FoMIa$ calculation,
    	and allow participants to optimize accordingly. 
	Also, to help check for buggy submissions, please 
	include your evaluation of the Ia-purity and Ia-efficiency
    	for the \specy\ confirmed subset.
\end{itemize}

\bigskip
% ==============================================================
% BIBLIOGRAPHY

\bibliographystyle{apj}
\bibliography{SNchallenge}

% ####################################
  \end{document}